\documentclass[11pt]{article}
\usepackage{amsmath,amssymb}
\usepackage{hyperref}
\usepackage{graphicx}

\newcommand{\bea}{\begin{eqnarray}}
\newcommand{\eea}{\end{eqnarray}}
\newcommand{\beq}{\begin{equation}}
\newcommand{\eeq}{\end{equation}}
\renewcommand{\Im}{\operatorname{Im}}
\renewcommand{\Re}{\operatorname{Re}}

\newcommand{\pd}{\partial}

\newsavebox{\uuunit}
\sbox{\uuunit}
    {\setlength{\unitlength}{0.825em}
     \begin{picture}(0.6,0.7)
        \thinlines
        \put(0,0){\line(1,0){0.5}}
        \put(0.15,0){\line(0,1){0.7}}
        \put(0.35,0){\line(0,1){0.8}}
       \multiput(0.3,0.8)(-0.04,-0.02){10}{\rule{0.5pt}{0.5pt}}
     \end {picture}}

\def\beq{\begin{equation}}
\def\bee{\begin{equation}}
\def\eeq{\end{equation}}
\def\bea{\begin{eqnarray}}
\def\eea{\end{eqnarray}}
\def\bd{\begin{displaymath}}
\def\ed{\end{displaymath}}

\numberwithin{equation}{section}

\begin{document}

\thispagestyle{empty}
{}

\hfill WITS-CTP-136 \\[-3ex]

\vskip22mm

\vskip -3mm
\begin{center}
{\bf\LARGE
\vskip - 1cm
Heating up branes in 
  gauged supergravity \\ [2mm]
}

\vspace{10mm}

{\large

{\bf Kevin Goldstein,} 
{\bf Suresh Nampuri,}
{\bf \'Alvaro V\'eliz-Osorio}

\vspace{1cm}

{\it
NITheP, School of Physics and Centre for Theoretical Physics \\ [1mm]
 University of the Witwatersrand, Johannesburg \\ [1mm] 
WITS 2050, South Africa\\ [5mm] 
}}
email: Kevin.Goldstein@wits.ac.za, nampuri@gmail.com, aveliz@gmail.com

\vspace{5mm}

\end{center}
\vspace{5mm}

\begin{center}
{\bf ABSTRACT}\\
\end{center}

In this note, we explore the solution space of non-extremal black objects in $4D$
and $5D$ ${\cal N}=2$ gauged supergravity in the presence of fluxes. We
present first order rewritings of the $4D$ action for a classes of
non-extremal dyonic and electric solutions with electric flux
backgrounds. Additionally, we obtain the non-extremal version of the Nernst brane
in $AdS_5$ using a simple deformation. Finally, we develop a new technique to deform extremal black
solutions in $4D$ to non-extremal solutions by an analysis of the
symmetries of the equations of motion.

\clearpage
\setcounter{page}{1}

\tableofcontents

\section{Introduction}
Black holes offer excellent laboratories for testing the predictions
of any purported quantum theory of gravity such as string theory. In
particular, one of the prime objectives of any such theory is to
provide a microscopic understanding of the Hilbert space of these
objects and thereby arrive at a first principle derivation of their
thermodynamic properties which can be written down heuristically in
Einstein's gravity. Furthermore, under gauge-gravity duality, black
objects are presumably  holographically dual to thermal ensembles in a field
theory. If this is the case, at strong coupling, field theoretic phenomena in the presence of such an
thermal ensemble, can be modeled in terms of the physics of
matter in the black background. In particular, black solutions which obey the third law of
thermodynamics, having vanishing entropy at zero temperature (Nernst
solutions), could be very useful in understanding quantum critical phase
transitions in the dual field theory. Hence, developing and exploring
the solution spaces of black objects in actions arising as low-energy limits of string theory -- namely certain supergravities 
 (SUGRA) -- is of paramount import in understanding non-perturbative
aspects of string theory and potentially, strongly coupled field
theories.
\par
The equations of motion (EOMs) of these supergravity actions are
second-order non-linear and hence enormously non-trivial to solve. For
a special class of extremal black solutions in $4D$ and $5D$ ${\cal N}=2$ gauged
supergravity, considerable advances have been made in the last
half-decade (see cf. \cite{5d}, \cite{4d}, \cite{giang} and
\cite{denef}). These solutions are amenable to a first order 
sum-of-squared-terms rewriting, which simplifies the problem to solving first order equations.
A natural generalization 
is  to develop techniques to study non-extremal black
solutions. In particular, given the fact that a large class of extremal solutions are known, we investigate methods
of generating non-extremal solutions from extremal ones which could be of great practical value. 
\par
Specifically, after a brief review of the action and the $4 D$ EOMs in
section \ref{section1}, we develop two first order first-order
rewritings of this action for a class of non-extremal electric and
dyonic black states with electric fluxes, in section
\ref{section3}. In section \ref{section4}, we deform the parameter
space of a class of extremal black solutions in asymptotically $AdS_5$
spaces ($AAdS_5$) to obtain the non-extremal equivalent of a
zero-entropy black Nernst brane in the same space. In 
section \ref{sec;hot} , we analyze the second order EOMs in $4D$ to identify certain
symmetries. We exploit these symmetries to compensate non-extremal deformations of the metric with
transformations of the parameters of the solution. The techniques developed in each section are illustrated by
generating example black solutions.  Our formal results
are applicable to black solutions with flat, spherical or hyperbolic
horizons. Finally in the conclusion we discuss possible extensions and
applications to $5D$.

\section{Equations of motion and extremal black objects}\label{section1}
The main purpose of this note is to study black gravitational
solutions of ${\cal N}=2$ gauged supergravity. We consider Lagrangians
of the form 
\begin{equation}
\label{action1} 
S= \int dx^4\sqrt{-g}
\left(
  R-2g_{i\bar\jmath}\pd_\mu z^i\pd^\mu \bar
  z^j-f_{IJ}F^I_{\;\mu\nu}F^{J\mu\nu}-\frac{1}{2}\tilde f_{IJ}
  F^I_{\;\mu\nu} F^J_{\;\rho\sigma}\epsilon^{\mu\nu\rho\sigma}-2 V_{g}
  (z,\bar z)
\right),   
\end{equation}
where the $z^i$ are complex scalar fields, and the $F^I$ are abelian gauge fields.
The structure of the different pieces of the
Lagrangian -- namely $f_{IJ}$, $g_{i\bar\jmath}$ and $ V_{g} (z,\bar z)$ -- and the range of the indices ($i$, $I$),  can
be found from very specific prescriptions (see Appendix
\ref{appendix:notation}), but for the sake of generality, we make
no assumptions at present.

For the action (\ref{action1}), Einstein's equations read
\beq\label{einstein}
   R_{\mu\nu}-2g_{i\bar\jmath}\pd_\mu z^i\pd_\nu \bar z^j= f_{IJ}\left(2 F^I_{\;\mu\lambda}F_{\;\nu}^{J\;\lambda}-\frac{1}{2}g_{\mu\nu}F^I_{\;\kappa\lambda}F^{J\kappa\lambda}\right)+g_{\mu\nu} V_g (z,\bar z).
\eeq 
As we are interested in studying static solutions with purely radial dependence, we take  $z^i=z^i(r)$, assume the line element is of the form
\beq\label{metric}
ds^2= -e^{2 \alpha(r)} dt^2+e^{-2 \alpha(r)}dr^2+ e^{2 \beta(r)}d\Omega^2_{k},
\eeq
with our ansatze for the gauge fields  given explicitly in Appendix \ref{sec:full-ansatz-black}. The label $k$ in (\ref{metric}) denotes  the assumed metric of the transverse spacial foliation with 
$k=-1,0,1$ standing for hyperbolic, flat and spherical respectively. We will also use the alternative notation 
\beq
  a(r)=e^{ \alpha(r)}\hspace{9mm} b(r)=e^{ \beta(r)}\hspace{9mm}  a(r)b(r)=e^{ \psi(r)}
\eeq
for the warp factors whenever this makes the expressions more compact.

 Given the assumptions mentioned above, Einstein's equations become
\beq\label{e1}
\left(a^2b^2\right)''= 2\left(k-2b^2V_g\right) ,
\eeq
and
\beq\label{e2}
 \frac{b''}{b}=-g_{i\bar\jmath}\pd_r z^i \pd_r \bar z^j,
\eeq
where ``$\;'\;$'' denotes derivatives with respect to $r$.
In addition to these, the following first-order constraint must be fulfilled 
\beq\label{e4}
 - \frac{1}{2}\left(a^2(b^2)'\right)'=\frac{V_{b}}{b^2}+b^2V_g-k.
\eeq
 In the previous expressions we introduced the  potential
 \begin{equation}
   \label{eq:Vb}
   V_{b}= f^{IJ}\left(Q_I-\tilde f_{IK} P^K\right)\left(Q_J-\tilde f_{JK} P^K\right)+f_{IJ}P^IP^J, 
 \end{equation}
with $f^{IJ}$ the inverse of $f_{JK}$.
This potential encodes all the contributions coming from the gauge field's electric $Q_I$ and magnetic $P^I$ charges (see Appendix \ref{sec:full-ansatz-black}).
In addition to (\ref{e1}), (\ref{e2}) and (\ref{e4}) we have the equations of motion for the scalars:
\beq\label{e3}
\pd_r\left(a^2b^2 g_{k\bar\jmath}  \pd_r \bar z^j\right) =a^2b^2\pd_k\left( g_{i\bar\jmath}\right)\pd_r z^i \pd_r \bar z^j+ \frac{\pd_kV_{b}}{b^2}+b^2\pd_kV_g.
\eeq
Clearly, it is rather complicated to solve the equations above in general. There exist, however, powerful techniques that are applicable for 
certain subclasses of solutions such as extremal black solutions and BPS black states. In the following we will concentrate mainly on solutions 
which are asymptotically $AdS_4$.\footnote{These solutions are specially relevant for $AdS_4/CFT_3$ holography.}

Let us start by discussing extremal  black solutions in $AdS_4$, these are solutions whose near-horizon geometry is described by an $AdS_2\times \Sigma_k$ metric, where $ \Sigma_1=S^2$ and  $ \Sigma_0=\mathbb{R}^2$. 
On this kind of backgrounds, $b(r)$ takes up a constant value, denoted by $\sigma$, encoding the entropy of the solution while 
for the other warp factor we have
\beq
  a=\frac{r}{l_{2}},
\eeq
where $l_2$ corresponds to the $AdS_2$ radius. Inserting this geometry into (\ref{e2}) and (\ref{e4}) we find that the scalars become constants whose values are determined by
\beq\label{att1}
 \partial_k V_b =-\sigma^4 \partial_k V_g. 
\eeq
Therefore they depend only on the values of the fluxes and charges of the system.\footnote{This is the well-known attractor mechanism (see cf. \cite{Ferrara}, \cite{Goldstein1, Goldstein12})}
Meanwhile, the two parameters that specify the metric can be extracted from the algebraic relations 
\beq
k= \frac{V_b}{\sigma^2} +\sigma^2 V_g\hspace{8mm} \left(\frac{\sigma}{l_2}\right)^2= \frac{V_b}{\sigma^2} -\sigma^2 V_g,
\eeq
derived from (\ref{e1}) and (\ref{e2}). For the case of black branes, i.e. solutions with 
flat horizon topology ($k=0$) the equations (\ref{att1}) yield the simple relations 
\beq
l_2^2=-\frac{1}{2V_g}\hspace{8mm} \sigma^4=-\frac{V_b}{V_g}, 
\eeq
which relate the geometric parameters to the potentials in a transparent way.

We turn our attention now to the asymptotic behavior. We want the metric to approach $AdS_4$ at ifinity i.e.
\beq
  a\xrightarrow{\hspace{0.2mm}r\rightarrow\infty\hspace{0.2mm}}b\xrightarrow{\hspace{0.2mm}r\rightarrow\infty\hspace{0.2mm}}\frac{r}{l_{4}}.
\eeq
Once again by inserting this into the equations of motion, we find that the scalars take up constant values which now
are determined by 
\beq\label{att2}
 \pd_k V_g=0,
\eeq
in other words, the scalar fields must flow towards critical points of the $V_g$ potential at infinity.
While the only geometric parameter, the $AdS$ radius, reads 
\beq
  l_4^{\,2} = -\frac{3}{V_g}.
\eeq

Therefore, if for certain given quantum numbers we wish to find an extremal black hole/brane, first we need to show that both  
(\ref{att1}) and (\ref{att2}) have compatible solutions that give rise to sensible physical parameters (e.g. positive $AdS$-radii) and 
then we must find solutions of the equations of motion $a(r)$, $b(r)$ and $z^i(r)$ that interpolate between the two $AdS$ geometries.

The procedure described above is in general challenging,  but in the context of ${\cal N}=2$ gauged supergravity,  there is a large class of extremal black branes that can be described by 
a simpler set of first order equations equations \cite{giang,4d}. As a matter of fact, interpolating solutions corresponding to these
equations are characterized by the relation
\beq
 a^2b^2=r \left(\frac{r}{l^{\,2}_4} +\frac{\sigma}{l_2}\right), 
\eeq
with the scalar field solutions  constructed as combinations of harmonic functions (\cite{steph}, \cite{Nick}).

\section{First-order flow equations for non-extremal branes}\label{section3}

As mentioned at the end of the previous section, being able to provide
a first order description for the system of interest makes the problem
of finding solutions much more tractable. Given that for ${\cal N}=2$
gauged supergravity models, extremal configurations can be described by
first-order equations, it is natural to wonder whether non-extremal
configurations can be found in a similar fashion. Fortunately it is in fact possible to find first order
descriptions for non-extremal configurations (\cite{4d},\cite{5d},
\cite{Gnecchi} and \cite{Cardoso:2008gm}). Unfortunately, these descriptions are
generically non-trivial to construct and can be set up only for
certain subclasses of black objects. In the following, we construct two new
first-order rewritings for non-extremal black branes: one for
generating classes of electric black branes and the other for dyonic
solutions.
\subsection{Electric black brane rewriting}
Consider the line element,
\begin{equation}
dx^2=-e^{2\alpha (r) }dt^2+e^{2\nu(r) }dr^2+e^{2\beta(r) }\left(dx^2+dy^2\right) \;.
\label{staticbb}
\end{equation}
The first rewriting that we find in this section is valid for solutions supported by electric charges and fluxes. Inserting the line element
(\ref{staticbb}) and our gauge ansatz (Appendix \ref{sec:full-ansatz-black}) into the $4D$  ${\cal N}=2$ gauged supergravity action (\ref{action1}), yields
the one-dimensional effective lagrangian density
\begin{eqnarray}
{\cal L}=e^{2\beta+\alpha-\nu}\left(\left(\beta '\right)^2+2\beta ' \alpha'-N_{IJ}\left(X^I\right)' \left(\bar X^J\right)'
- e^{2\nu} \, V_{\rm tot} ( X,\bar X) \right) \;,
\end{eqnarray}
where 
\beq
   V_{\rm tot}= g^2 V_g+e^{-4\beta} V_b,
\eeq
$N_{IJ}$ is given by (\ref{eq:N-big}) and the scalars $X$ are related to the scalars $z$ by (\ref{eq:zX_relation},\ref{D-gauge})
Here, the constant $g^2$ is inserted to keep track of terms arising from the flux potential, $V_g$.

As a first step, we decompose the warp factors $\alpha$ and $\nu$ appearing in the line-element \eqref{staticbb} into
\begin{equation}
\alpha=\alpha_0+\frac{1}{2} \ln f\;\;\; \;\;\; \nu=\nu_0-\frac{1}{2} \ln f \,,
\end{equation}
where                                 
\begin{equation}
 f=-\mu \, r+g^2e^{2\alpha_1(r)}.
\end{equation}
The horizon will correspond to physical solutions of the equation 
\beq\label{horizon}
  \mu \, r = g^2e^{2\alpha_1(r)}.
\eeq
As we will see in some examples later, by changing the value of $\mu$ we modify the root structure of this equation so that
the constant parameter $\mu$ parametrizes the non-extremality of the solution. 
 To proceed, we decompose ${\cal L}$ into powers
of $g^2$, 
\begin{equation}
 {\cal L}={\cal L}_0+g^2{\cal L}_2.
\end{equation}

The next step is to rewrite both ${\cal L}_0$ and ${\cal L}_2$ in terms of squares of first-order flow equations. 
From this procedure, we will obtain an overdetermined system whose consistency must be checked.
This rewriting is analogous to the one performed in five-dimensional ${\cal N}=2$, $U(1)$ gauged supergravity in 
\cite{Cardoso:2008gm,5d}.
Let us start with ${\cal L}_0$. For the sake of simplicity, we use the diffeomorphism invariance of the theory, changing the radial coordinate, to set 
$\nu_0=2\beta+\alpha_0$  and obtain
\begin{equation}
{\cal L}_0 =-\mu \beta'-\mu r\left(\left(\beta '\right)^2+2\beta' \alpha_0'-N_{IJ}\left(X^I\right)' \left(\bar X^J\right)'\right)- 
e^{2\alpha_0}{\cal G}^{IJ}Q_I Q_J. 
\end{equation}
Next, we introduce the combination
\begin{equation}
{\cal E}^I = X'^I - \alpha_0' X^I + e^{\alpha_0} N^{IJ} \lambda_J \;,
\end{equation}
where $\lambda_J$ are real parameters that are  a priori arbitrary. This expression is the candidate for the first order equation driving the scalar fields.  Using \eqref{D-gauge} we get
\begin{equation}
{\bar X}^I N_{IJ} {\cal E}^J = {\bar X}^I N_{IJ} X'^J + \alpha_0' + e^{\alpha_0} \lambda_I {\bar X}^I \;.
\end{equation}
Inserting the above equation into ${\cal L}_0$ we find
\begin{eqnarray}
{\cal L}_0 &=& -\mu \beta' - \mu \lambda_I \left[ r e^{\alpha_0} \left( X^I + {\bar X}^I \right)\right]' \nonumber\\ 
&&
-\mu r \left[ \left(\beta'+ \alpha_0'\right)^2 -{\cal G}^{IJ}
N_{IP} {\cal E}^P   N_{JQ} \bar{\cal E}^Q  + 2 \, {\bar X}^I N_{IJ} {\cal E}^J \,  X^K N_{KL} \bar {\cal E}^L
\right. \nonumber\\
&&  \qquad \quad \left. - 2 \left( \alpha_0' + e^{\alpha_0} \lambda_I X^I \right) \left( \alpha_0' + e^{\alpha_0} \lambda_I {\bar X}^I \right) 
 \right] \nonumber\\
&& + \mu e^{\alpha_0} \lambda_I \left( X^I + {\bar X}^I \right) - e^{2 \alpha_0} {\cal G}^{IJ} \left(r  \mu \, \lambda_I \lambda_J + Q_I Q_J \right) 
\;,
\label{L0-BPS}
\end{eqnarray}
where we introduced the positive definite matrix 
\begin{equation} \label{GIJ}
{\cal G}^{IJ}= N^{IJ} + 2 e^{{K}}\, \bar{{ X}}^I { X}^J \;,
\end{equation}
defined on the scalar moduli space with the K\"ahler potential, $K$, given by~(\ref{eq:Kahler}).
Notice that the first line  of \eqref{L0-BPS}
contains total derivatives, while the second and third lines contain squares of various combinations of fields.
The last line encodes additional constraints that will be discussed soon.

Requiring the variation of the second and third lines of ${\cal L}_0$
with respect to the fields ($\beta , \alpha_0, X^I$) to vanish can be
achieved by setting the various squares to zero.  This yields $X^I =
{\bar X}^I$ as well as the first-order flow equations
\begin{eqnarray}
\beta' \hspace{-2mm}&=&\hspace{-2mm} - \alpha_0' \;, \nonumber\\
{\cal E}^I\hspace{-2mm} &=&\hspace{-2mm} 0 \;, \nonumber\\
\alpha_0'\hspace{-2mm} &=&\hspace{-2mm} -  e^{\alpha_0} \lambda_I X^I \;.
\label{first-order}
\end{eqnarray}
These can be readily integrated by introducing the rescaled scalar fields 
\begin{equation}
Y^I = e^{- \alpha_0} X^I \;,
\end{equation}
giving
\begin{eqnarray}
e^{2\beta}\hspace{-2mm} &=&\hspace{-2mm} e^{-2\alpha_0} = Y^I \, H_I \;, \nonumber\\
Y^I\hspace{-2mm} &=&\hspace{-2mm} - N^{IJ} H_J \;, 
\label{first-sol}
\end{eqnarray}
where $H_I = \gamma_I + \lambda_I \, r$, with constant $\gamma_I$.  To obtain (\ref{first-sol}) we used the relations
$F_{IJK} X^K =0$ (\ref{eq:homog-rel}) and
$\left( X^I N_{IJ} {\bar X}^J \right)' =0$ (\ref{D-gauge}), as well as the fact that the $X^I$'s are real.

To conclude the ${\cal L}_0$ rewriting, we must analyze the variation
of the terms in the last line of \eqref{L0-BPS}. Their variation
should result in equations that are consistent with the flow equations
\eqref{first-order}.  To this end, we restrict the analysis to models
in which satisfy two simplifying conditions.  First, we demand that there are no
linear terms proportional to $(X - \bar X)^K$ in ${\cal G}^{IJ}$, i.e.
\begin{equation}
{\cal G}^{IJ} (X, \bar X) = {\cal G}^{IJ} ({\rm Re} X, {\rm Re} X) + {\cal O} \left( (X - \bar X)^2 \right) \;,
\end{equation}
so that we may solve the variational equations steming from the last line of 
\eqref{L0-BPS} by setting $(X - \bar X)^I=0$.  Second, we assume that 
the 
physical charges $Q_I$ can be expressed in terms of the constant vectors $\lambda_I$ and $\gamma_I$ by using the relation
\begin{equation}
{\cal G}^{IJ} ({\rm Re} X, {\rm Re} X) \, Q_I Q_J = \mu {\cal G}^{IJ} ({\rm Re} X, {\rm Re} X)  \, \lambda_I \gamma_J \;.
\label{rel-Q-qg}
\end{equation}
Generically, this can be achieved in models in which ${\cal G}^{IJ} ({\rm Re} X, {\rm Re} X) $ is diagonal or off-diagonal.  

Imposing the relation $(X - \bar X)^I=0$ as well as \eqref{rel-Q-qg}, the last line of \eqref{L0-BPS} becomes
\begin{equation}
\mu e^{\alpha_0} \left( 2 \lambda_I x^I  - 
e^{\alpha_0} {\cal G}^{IJ}(x,x) \,  H_I \lambda_J \right)
\;,
\label{Delta}
\end{equation}
where $x^I = {\rm Re} X^I$.
Denoting the combination \eqref{Delta} by $\Delta$,
we compute its variation with respect to $\alpha_0$ and with respect to the scalar fields $x^I$, 
respectively, and obtain
\begin{eqnarray}
\label{var-delta}
&& \mu \, \delta \left( e^{\alpha_0} \right) \left[ 2 \lambda_I x^I  - 
e^{\alpha_0} {\cal G}^{IJ}(x, x)  \, H_I \lambda_J 
\right] =0 \;, \\
&& \mu \, e^{\alpha_0} \, 
\delta x^L \left[ 2 \lambda_L - i 
e^{\alpha_0} N^{IP} \left( F_{PQL} - {\bar F}_{\bar P \bar Q \bar L} \right) N^{QJ} H_I \lambda_J - 2 e^{\alpha_0}
 \left(\left( {x}^I H_I \right) \lambda_L  + 
 \left({x}^I \lambda_I \right) H_L  \right) 
\right] 
=0 \;, \nonumber
\end{eqnarray}
where in the second equation the quantities $N^{IJ}$ and $F_{IJK}$ are evaluated at $x^L$.
Using \eqref{first-sol}, we find that the first equation  of 
\eqref{var-delta} is satisfied, while for the second equation we obtain
\begin{equation}
\mu \, e^{\alpha_0} \,  \left({x}^I \lambda_I \right) 
\delta x^L \, N_{LK} \, x^K  =0 \;,
\label{var-Delta}
   \end{equation}
Recalling that the variation of \eqref{D-gauge} yields $\delta X^I \, N_{IJ} {\bar X}^J + 
X^I N_{IJ} \delta  {\bar X}^J  =0$, and setting  $X^I = {\bar X}^I$, gives
$\delta x^I  \, N_{IJ} \,  x^J =0$.  This equals \eqref{var-Delta},
 hence we conclude that for a model satisfying \eqref{rel-Q-qg} the conditions 
\eqref{var-delta} stemming from the variation of $\Delta$ are satisfied by a solution
to the first-order flow equations \eqref{first-order}.

Now we turn our attention to ${\cal L}_2$, which is given by 
\begin{equation} \label{L2}
 {\cal L}_2=e^{2\alpha_1}\left(\left(\beta'\right)^2+2\beta'\left(\alpha_0'+\alpha_1'\right)-N_{IJ}\left(X^I\right)' \left(\bar X^J\right)'-e^{4\beta+2\alpha_0-2\alpha_1}V_g(X, \bar X)\right).
\end{equation}
Following \cite{5d}, we rewrite this as
\begin{eqnarray} \label{L2_b}
 {\cal L}_2 &=&e^{2\alpha_1}\left\{\left(\beta' + \alpha_0' + \alpha_1' + 2 \alpha \, e^{2A + \alpha_0 - \alpha_1} \, h_I \, X^I \right)
 \left(\beta' + \alpha_0' + \alpha_1' + 2 \alpha \, e^{2\beta + \alpha_0 - \alpha_1} \, h_I \, {\bar X}^I \right) \right. \nonumber\\
 && \left.
 - \left(\alpha_0' + \alpha_1' + \alpha \, e^{2\beta + \alpha_0 - \alpha_1} \, h_I \, X^I \right)
 \left(\alpha_0' + \alpha_1' + \alpha \, e^{2\beta + \alpha_0 - \alpha_1} \, h_I \, {\bar X}^I \right) \right. \nonumber\\
 && \left.
- \left(\beta'  + \alpha \, e^{2\beta + \alpha_0 - \alpha_1} \, h_I \, X^I \right)
 \left(\beta'  + \alpha \, e^{2\beta + \alpha_0 - \alpha_1} \, h_I \, {\bar X}^I \right) \right. \nonumber\\
 && \left.
  -N_{IJ}\left((X^I)' + \beta' X^I   - \kappa \, e^{2\beta + \alpha_0 - \alpha_1} \, N^{IK} h_K \right)
 \left( ({\bar X}^J)'
 + \beta' {\bar X}^J   - \kappa \, e^{2\beta + \alpha_0 - \alpha_1} \, N^{JL} h_L \right)
\right\} \nonumber\\
&& - \kappa \left[ e^{2\beta + \alpha_0 + \alpha_1} \, h_I \, \left( X^I + {\bar X}^I \right) \right]' \;,
\end{eqnarray}
where $\kappa = \pm 1$ (corresponding to two possible rewritings). This expresses ${\cal L}_2$ in terms of squares of combinations of fields.
Setting these squares to zero ensures the vanishing of the variation of 
${\cal L}_2$ with respect to the various fields.
This yields
$X^I = {\bar X}^I$ as well as the first-order flow equations
\begin{eqnarray}
\beta' + \alpha_0' + \alpha_1' &=& - 2 \kappa \, e^{2\beta + \alpha_0 - \alpha_1} \, h_I \, X^I \;, \nonumber\\
\alpha_0' + \kappa_1' &=& -  \kappa \, e^{2\beta + \alpha_0 - \alpha_1} \, h_I \, X^I \;, \nonumber\\
\beta'  &=& -  \kappa \, e^{2\beta + \alpha_0 - \alpha_1} \, h_I \, X^I \;. \nonumber\\
(X^I)' + \beta' X^I   &=& \kappa \, e^{2\beta + \alpha_0 - \alpha_1} \, N^{IK} h_K \;.
\end{eqnarray}\footnote{A first order rewriting for only non-extremal black holes, with a different reparametrization  was independently developed in \cite{Gnecchi}.}
By demanding compatibility with the first-order flow equations \eqref{first-order} obtained from ${\cal L}_0$ 
we find 
\beq
2\alpha_0 + \alpha_1 =0
\eeq
 as well as 
\beq
\lambda_I = - \kappa \, h_I.
\eeq
 The harmonic functions $H_I$ entering in the solution
\eqref{first-sol} are thus given by 
\beq
H_I = \gamma_I - \kappa \, h_I \, r.
\eeq  
Summarizing, we have obtained a non-extremal static electrically charged black brane solutions are described by 
\begin{eqnarray}
ds^2\hspace{-2mm} &=& \hspace{-2mm}-e^{-2\beta} \, f \, dt^2+e^{2 \beta} \, f^{-1} \, dr^2+e^{2\beta}\left(dx^2+dy^2\right) \;,
\nonumber\\
f\hspace{-2mm} &=&\hspace{-2mm} - \mu \, r + g^2 \, e^{4 \beta} \;, \nonumber\\
e^{2\beta}\hspace{-2mm} &=&\hspace{-2mm} Y^I \, H_I \;, \nonumber\\
Y^I \hspace{-2mm}&=& \hspace{-2mm}{\bar Y}^I =  - N^{IJ} H_J \;, \nonumber\\
H_I\hspace{-2mm} &=&\hspace{-2mm} \gamma_I - \kappa \, h_I \, r \;\;\;,\;\;\; \kappa = \pm 1 \;,
\label{first-order-sol}
\end{eqnarray}
where the constants $\gamma_I$ are related to the physical charges $Q_I$ by 
\beq{\cal G}^{IJ} Q_I Q_J = - \kappa \, \mu  {\cal G}^{IJ} \, h_I \gamma_J.
\eeq

In the following we discuss two examples. First, we consider an ${\cal N}=2$ model based on the prepotential
\beq\label{prepotential}
F(X) = - 2i \sqrt{X^0 X^1 X^2 X^3}.
\eeq
This model comes from the $U(1)^4$ truncation of ${\cal N }= 8$  gauged supergravity in four dimensions \cite{herm}.
Using (\ref{NIJ}), we compute $N_{IJ}$ for this prepotential using  and find
\begin{equation}
N_{IJ} = \frac{i}{2} \,  \left( 2 \delta_{IJ} -1 \right) \frac{F(X)}{X^I X^J}\;.
\label{NIJ}
\end{equation}
Hence (\ref{GIJ}) reads
\begin{equation}
 {\cal G}^{IJ}=\frac{1}{4}X^IX^J\left(8 + (-1)^{\delta_{IJ}}\frac{2i}{F(X)}\right).
\end{equation}
Inserting \eqref{NIJ} into \eqref{D-gauge} yields
\begin{equation}
F(X) = - \frac{i}{4} \;,
\end{equation}
therefore
\begin{equation}
 {\cal G}^{IJ}=4\left(X^I\right)^2\delta^{IJ} \;,
\end{equation}
as well as 
\begin{equation}
e^{2\beta} = 4 i \, F(Y) = 8 \sqrt{Y^0 Y^1 Y^2 Y^3} \;.
\label{A-Y}
\end{equation}
Using \eqref{NIJ} we compute
\begin{equation}
N_{IJ} \, Y^J = - i \, \frac{F(Y)}{Y^I} \;,
\end{equation}
and hence
\begin{equation}
Y^I = \frac14 \, \frac{e^{2\beta}}{H_I} \;,
\end{equation}
where we made use of the third equation of \eqref{first-order-sol}.
Explicitly, the harmonic functions are given by
\beq
 H_I= \frac{1}{ h_I}( \mu^{-1}\,Q_I^2+h^2_I r).
\eeq
  Inserting this into \eqref{A-Y} gives the warp factor
\begin{equation}
e^{2\beta} = 2 \sqrt{H_0 H_1 H_2 H_3} \;.
\end{equation}
These solutions correspond to the brane solutions in \cite{sabra}. They interpolate between an $AdS_4$ asymptopia with 
\beq
  l_4^2\sim \frac{1}{\sqrt{h_0h_1 h_2h_3}}
\eeq
and the outer horizon given by the largest root of (\ref{horizon}). It is possible to tune $\mu$ to make the solution extremal 
by demanding that the following holds at horizon
\beq
  -\frac{\mu}{4}=g^2\beta' e^{4\beta}.
\eeq

As a second example consider the model based on the prepotential 
\beq
F(X) = - X^0 X^1.
\eeq
  For this prepotential we have 
\begin{equation}
N_{IJ} =-2|\epsilon_{IJ}|
\label{NIJ2}
\end{equation}
For axion-free configurations, (\ref{GIJ}) is simply 
\begin{equation}
 {\cal G}^{IJ}=\frac{1}{2}\left( \begin{array}{cc}
z^{-1} & 0 \\
0 &  z  \end{array} \right),
\end{equation}
where we introduced 
\begin{equation}
  z= \frac{X^1}{X^0}.
\end{equation}
Using \eqref{D-gauge}, we find 
\begin{equation}
e^{2\beta} = 4\, Y^0 Y^1.
\end{equation}
While the scalars are given by 
\begin{equation}
 Y^J =  \, \frac{1}{2} |\epsilon^{IJ}| H_J \;.
\end{equation}
Once again, we have a solution that interpolates between $AdS_4$ with radius
\beq
    l_4^2\sim \frac{1}{|h_0h_1|}
\eeq
and the corresponding horizon.

\subsection{Dyonic rewriting}
Here we develop a first order rewriting to generate a class of dyonic non-extremal solutions in the presence of electric fluxes with $h_0 = P^0 =0$.
For this set of quantum numbers, the 1D Lagrangian density can be rewritten as a sum of perfect squares which yield the following first order equations:
\beq
(z^i )'=\frac{1}{2} g^{i j}( \gamma_j -2 r\, \tilde q_j)e^{-\frac{8}{3}\beta}
\eeq
\beq
 \beta'=\left(-\frac{1}{4}z^ih_i+\frac{1}{2} g_{ij} z^i P^j\right) e^{-\alpha_1}
\eeq
\beq
  \alpha_ 1'= 2\beta'+2\,Q_0 e^{-2\beta-\alpha_1}
\eeq
with constraints given by :

\beq
  h_iP^i=0\hspace{8mm} -2Q_0=\gamma_i\,P^i\hspace{8mm}g^{ij} q_i q_j =- 4\mu g^{ij} \tilde q_i\gamma_j
\eeq
\beq
 \frac{1}{3} e^{-2 \beta}z^i\tilde q_i+  \frac{1}{3}z^ih_i e^{-\alpha_1}-\frac{2}{3}g_{ij} z_i P^j e^{-\frac{4}{3}\beta-\alpha_1}
\eeq
Notable solutions of the above equations for $Q_0=0$ contain the
$\eta$-geometries discussed in \cite{Donos} with relevance to
holographic condensed matter issues.  Solutions of these equations can
be up-lifted to $5D$ and be shown to be solutions of the first order
equations derived from a re-writing of the 5D gauged SUGRA action for
the same set of quantum numbers, written in $5D$ language.

\section{Hot Nernst brane in $AdS_5$}
\label{section4}
In this section we use first order rewritings for ${\cal N}=2$ 5D gauged supergravity \cite{5d} to discover and write down a new class of black solutions in 
 $AAdS_5$ spaces which
in the extremal limit have vanishing entropy. In the extremal limit these satisfy the third law of thermodynamics or the Nernst law and hence, go by the name of Nernst branes. Studying the physics of
matter in near-extremal solutions of the Nernst  type will potentially shed
valuable light on the physics of phase transitions at zero-temperature
in the dual $4D$ field theory, and the phase diagram that captures
aspects of this transition as one approaches absolute zero. We will
look for these solutions by extremizing the low-energy $5D$ ${\cal
  N}=2$ gauged supergravity action arising from the compactification
of M-theory on a $CY_3$. This action is amenable to first order
rewritings for dyonic black solutions \cite{5d} in the presence
of electric fluxes and which are charged under one of
the Cartans of the angular momentum group in $5D$.  One can solve the resulting first order equations
for a system with no dyonic charges and only an angular momentum $J$
and with electric fluxes $h_1, h_2$ and $h_3$ in the STU model to
obtain a family of solutions (see Appendix C in \cite{5d}) wherein the
scalars, $X^A$, ($A=1,2,3$), are constant through out and the metric
is given by
\begin{equation}
ds^2= - e^{2U} dt^2 + \frac{d\tau^2}{\tau^2} + e^{2B}( dx^2 + dy^2) + e^{2W} (dz + C dt)^2
\end{equation}
and where 
\begin{eqnarray}
e^{B} &=& \tau \\
e^{2W}&=& (\alpha + \gamma \tau^4)/\tau^2\\
e^{2U}&=& \tau^6/(\alpha + \gamma \tau^4)\\
 C&=& \tau^4/(\alpha + \gamma \tau^4)
\end{eqnarray}
Here $\alpha = \frac{3 J}{4 h_A X^A}$ is a positive constant, while $\gamma$ is a constant of integration.
The above solution is a zero-entropy $AAdS_5$ extremal Nernst solution, which can be compactified to $4D$ and lies in the BPS class of solutions of the $4D$ rewriting, as observed in \cite{5d}.
\par
The Nernst brane, being at zero entropy, has no discernible scale in its near horizon geometry. A hot Nernst brane on the other hand would have a natural scale, arising from the temperature. In order to heat up a zero-scale system to a finite scale one, we take a hint from the observation that 
massive  BTZ black holes in $AdS_3$ can be thought of as deformations of the angular momentum from static $M=0$ BTZ black holes. Hence we perform a deformation of the $g_{tz}$ term in the above solution via the shift:
\begin{eqnarray}
C &\rightarrow& C+ \lambda\,.
\end{eqnarray}
One can check that the above solution satisfies the EOMs. 

In particular, 
for $\lambda = -\gamma <0$, we observe that the solution has two horizons with the outer horizon given  by the zero of $g_{00} = -e^{2U} + e^{2W} C^2$ at $\tau = (\alpha |\lambda|)^{1/4}$, and where the temperature, computed by demanding smoothness of the Euclidean near-horizon Rindler metric is 
\beq
T=\frac{2\pi}{|\lambda| \alpha (\alpha \lambda)^{1/4}}.
\eeq
The black solutions so obtained, interpolate between the non-extremal Rindler geometry near the horizon of the black brane to $AdS_5$.
Hence, we have discovered here a family of hot black branes in AAdS which obeys the third law of thermodynamics at zero temperature.

\section{Hot deformations}\label{sec;hot}
So far, we have written down rewritings which allow us to obtain hot black solutions by solving first order equations. However, these rewritings have various constraints on charges and fluxes and it is not possible to always obtain a non-extremal solution for a generic set of charges and fluxes whose extremal counterparts are fully known. The Nernst brane is a case in point. So far, there exists no first order re-writing for its non-extremal counterpart. We wrote down the hot solution by performing a deformation on the extremal solution to obtain a new solution of the EOMs. \par
In this section, we generalize this principle by developing a deformation algorithm for a large class of extremal solutions, that will automatically generate the corresponding non-extremal solutions.

We begin with an outline of the working philosophy behind the deformation of extremal dyonic solutions in $4D$ ${\cal N}=2$ gauged SUGRA, in the presence of fluxes, to non-extremal solutions. In order to generate new solutions of the second order EOMs, we identify deformations that are symmetries of the EOMs.
The deformations are implemented in three steps. First, notice that the deformations  
 \begin{eqnarray}\label{symmetry}
 (ab)^2 &\rightarrow &(ab)^2 - \mu r \\ b&\rightarrow &b \label{symmetry}
\end{eqnarray}
form a symmetry of the equations of motion \eqref{e1} and \eqref{e2}. We can promote 
this transformation to a symmetry of the full set of equations of motion by considering a transformation for which the functional forms of $b$ and $z^i$ is left intact, while the 
relationships between the constants -- ie. charges/fluxes and parameters -- is left 
undetermined. Then we procede to impose  \eqref{e4} and \eqref{e3} to hopefully find  new algebraic 
relationships between the constants. We have thus constructed  a
well-defined solution generating technique, which is applicable to a large class of extremal solutions where the above operations can be defined. 
\par
The set of deformations outlined above, when applied to extremal configurations,
generally produces a non-extremal configuration.  It must be noted
that for all known solutions in the STU model, this deformation works
algebraically, in the sense that one can always define an operation in
parameter space that elevates the deformation to a symmetry of the
system and hence the deformation generates a new-solution. However, in
some cases, these solutions, though mathematically well-defined do not
correspond to real physical backgrounds as they fail the standard
tests of positivity of metric coefficients or finiteness of
4-derivative scalars constructed from the Riemann, Ricci tensor and
the Ricci scalar. As a final step one has to check that the solutions obtained are physical.
\par
Below, we give an illustrative example of how to implement the three-step deformation technique for a well-known magnetic extremal black brane \cite{Cacc} to produce a well defined non-extremal solution.
This  solution has a  line element of the form (\ref{metric}) based on the 
prepotential (\ref{prepotential}).
It is parametrized by 4 magnetic charges $P^I$ and four electric fluxes $h_I$. For extremal solution of this type, the hamiltonian constraint 
is given by \cite{giang}
\beq\label{HC}
P^Ih_I=0. 
\eeq
We can solve this constraint by taking 
\beq
 P^0=-\frac{3}{16  h_0}\hspace{8mm}  P^i=\frac{1}{16 h_i}.
\eeq
The warp factors for this solution read 
\beq
  b^2= 8\prod_{I}\sqrt{\alpha^I r+\beta^I}\hspace{10mm}a\,b=(\gamma r^2 +\delta),
\eeq
where
\beq
 \alpha^I=-\frac{\gamma}{4 h_I}\hspace{8mm}  \beta^I=P^I
\eeq
$\gamma$ and $\delta$ are real constants subject to the constraint 
\beq
\gamma\delta= -\frac{3}{16}.
\eeq
Moreover, in order to avoid a naked singularity we must have $\gamma>0$ and the horizon is located
at 
\beq
  r_H=\frac{\sqrt{3}}{4\,\gamma}.
\eeq
This solution can be found by making use of the first order equations \cite{giang, 4d} for the prepotential 
(\ref{prepotential}).
We now implement the three-step hot deformation:
\begin{itemize}
\item[\bf Step 1:] Deform the warp factors 
  \beq 
  (a\,b)^2 \rightarrow (a\,b)^2-\mu r, 
  \eeq
 keeping the functional forms of $b$ and $z^i$ fixed.
\item[\bf Step 2:] Insert the deformation into the remaining equations and read off the corresponding algebraic equations. For the case at hand, these can be solved and they imply the modified relationships
\beq
    P^0=-\frac{3}{16  h_0}\; \longrightarrow\;  P^0\left(1+\frac{48}{9}\mu \gamma\right)^{-1/2} =  -\frac{3}{16  h_0}
\eeq
\beq
 P^i=\frac{1}{16  h_i}\; \longrightarrow\;  P^i \left(1-16\mu \gamma\right)^{-1/2} =  \frac{1}{16  h_i}.
\eeq
\item[\bf Step 3:] We check that the solution is physical.
\end{itemize}

In summary, we have obtained a regular solution with
\beq
 b^2= 8\prod_{I}\sqrt{\alpha^I r+\beta^I}\hspace{10mm}
  (ab)^2= \left( \gamma r^2 +\delta\right)^2 -\mu r,
\eeq
where
\beq
 \alpha^I=-\frac{\gamma}{4 h_I},
\eeq
and
\beq
\beta^0=P^0 \left(1+\frac{48}{9}\mu \gamma\right)^{-1/2}\hspace{8mm}\beta^i = P^i \left(1-16\mu \gamma\right)^{-1/2}.
\eeq
This solution is non-extremal as it has two horizons, which for small $\mu$, are located at
\beq
  r_\pm=r_H\,\pm \frac{\sqrt{\mu}}{\,3^{1/4}}+\dots \,.
\eeq
This technique should prove to be of tremendous utility in generating and studying hot solutions of relevance to holographic condensed matter and fluid dynamics, in the future.

\section{Conclusions}
In this note we have developed new rewritings of ${\cal N}=2$ $4D$
gSUGRA actions for a class of dyonic non-extremal and electric
non-extremal solutions not contained in the dyonic class, obtained the
non-extremal version of the Nernst brane in $AAdS_5$ and introduced a
deformation technique applied on extremal solutions to generate new
non-extremal ones. These techniques will be particularly useful in
developing holographic black duals for computations in 3D CFTs dual to
$AdS_4$ and Lifschitz geometries which can be obtained by choosing the
flux quanta appropriately. A recent development in this field (see
cf. \cite{steph}) has resulted in the existence of closed form
solutions for the metric in asymptotically $AdS$ spaces and where the
scalars can be obtained by solving algebraic equations. The new
classes of extremal solutions so obtained can now be plugged into the
deformation toolkit we have introduced here to obtain whole new
classes of non-extremal solutions. More excitingly, one should be able
to easily extend this formalism to $5D$ gauged supergravity where
whole new extremal and non-extremal solution spaces could be uncovered
with exciting implications for $4D$ field theories. For example, based on past
and recent research \cite{goldstein,Myers,Myers1}
and ongoing work by some of the authors \cite{future}, one can write
down the affine parameter that characterizes the flow of scalar fields
from boundary to the horizon of extremal black solutions and these are
holographically dual to the c-function that parametrizes Wilsonian
flow in the corresponding field theory. Hence one can make statements
about the central charges of the UV and IR fixed points of the
Wilsonian flow from trivial calculations in the bulk. Hence, the
techniques developed here have easy extensions and applications
which can impact not just the study of black holes but also
holographically coded processes of field theories.

\appendix
\section*{Acknowledgments}
The authors are especially grateful to Gabriel Lopes Cardoso for invaluable discussion. 
We are also grafeful to
 Michele Cirafici,
Michael Haack, Vishnu Jejjala, Giuseppe Policastro, Chiara Toldo and
Jan Troost for stimulating conversations related to the work in this
note.  S.N. and A.V.O gratefully acknowledge support from the Center
for Theoretical Physics at the University of the
Witwaterstrand. S.N. is supported by a URC fellowship, an NRF grant
and by a SARCHI fellowship.  The work of A.V.O is supported in part by
a URC fellowship.  The work of KG is supported in part by the National
Research Foundation. Any opinion, findings and conclusions or
recommendations expressed in this material are those of the authors
and therefore the NRF do not accept any liability with regard thereto.

\section{Notation}\label{appendix:notation}

In this appendix we follow the notation found in \cite{4d}, which for convenience is summarized below, and relate it to the parameters which appear in (\ref{action1}).
The Lagrangian describing the couplings of ${\cal N}=2$ vector multiplets to ${\cal N}=2$ supergravity
is encoded by a  holomorphic function $F(X)$ called the prepotential, that depends on  
$n + 1$ complex scalar fields $X^I$ ($I = 0, \dots, n$).  
 $F(X)$ is homogeneous of degree two, i.e. $F(\lambda X) = \lambda^2 F(X)$,
from which leads to  the homogeneity properties
\begin{eqnarray}
&& F_I = F_{IJ} \, X^J \;, \nonumber\\
&& F_{IJK} \, X^K = 0 \;,
\label{eq:homog-rel}
\end{eqnarray}
where $F_I = \partial F(X)/\partial X^I \,,\,
F_{IJ} = \partial^2 F/\partial X^I \partial X^J$, etc. The $X^I$ are redundant 
while the physical scalar fields
\begin{equation}
  \label{eq:zX_relation}
 z^i = X^i/X^0 \qquad (i = 1, \dots, n)  
\end{equation}
 parametrize an $n$-dimensional complex
hypersurface.  The redundancy of the $X^I$ is encoded by a constraint on the symplectic vector
$(X^I, F_I (X))$:
\begin{equation}
i \left( {\bar X}^I \, F_I -  {\bar F}_I \, X^I \right) = 1 \;.
\label{eq:einstein-norm}
\end{equation}
This can be written as
\begin{equation}
\label{D-gauge}
- N_{IJ} \, X^I \, {\bar X}^{J} = 1 \;,
\end{equation}
where 
\begin{equation}
N_{IJ} = -i \left( F_{IJ} - {\bar F}_{IJ} \right) \;.
\label{eq:N-big}
\end{equation}
%
For ${\cal N}=2$ based models, the lagrangian (\ref{action1}) is constructed out of 
\begin{equation}
g_{i  \bar{\jmath}} = \frac{\partial^2 K(z, \bar z)}{\partial z^i \, \partial {\bar z}^j}, 
\label{eq:kaehler-metric}
\end{equation}
where the K\"ahler potential is given by
\begin{equation}
  \label{eq:Kahler}
  e^{-K}= i (\bar X^I F_I-  X^I \bar F_I)\;,
\end{equation}
as well as 
\beq
   f_{IJ}  =  \frac{1}{2} \Im ({\cal N}_{IJ})\hspace{9mm}   \tilde f_{IJ}  =  \frac{1}{2} \Re ({\cal N}_{IJ}),
\eeq
where
\begin{equation}
{\cal N}_{IJ} = {\bar F}_{IJ} + i \, \frac{N_{IK} \, X^K \, N_{JL} \, X^L}{X^M \, N_{MN} \, X^N} \;.
\end{equation}
and
\beq \label{flux potential}
V_g= N^{IJ} \hat{h}_I \bar{\hat{h}}_J  - 2 e^K \,|W|^2 
\;,
\end{equation}
where
\beq
\hat{h}_I = h_I - F_{IK} h^K  \hspace{8mm}  W = h^I F_I - h_I X^I \;,
\eeq
and  $(h^I , h_I)$ denote the magnetic and electric fluxes respectively -- see \cite{Samtleben:2008pe} for a nice review.

\section{Explicit gauge field ansatz} \label{sec:full-ansatz-black}

In this appendix we show our explicit gauge field ansatz for the three cases $k=-1,0,1$. To establish coordinates we take
\begin{equation}
  \label{eq:trans_metric}
  d\Omega_k^2=
  \left\{
    \begin{array}{ll}
       d\theta^2 + \sin^2\theta d\phi^2 & k=1 \\
       dx^2+dy^2 & k=0 \\
       d\theta^2 + \sinh^2\theta d\phi^2 & k=-1 \\
    \end{array}
  \right.
\end{equation}
In these coordinates ansatze for the gauge fields are (cf.\cite{convention,Nick})
\begin{equation}
  \label{eq:gauge_anz}
  A^I= \mathcal{Q}^I e^{-2\beta}dt - P^I
  \left\{
    \begin{array}{ll}
        \cos\theta d\phi\,  & k=1 \\
       x dy\, & k=0 \\
      \cosh\theta d\phi\,  & k=-1 \\
    \end{array}
  \right.
\end{equation}
where 
\begin{eqnarray}
\mathcal{Q}^I = f^{IJ}(Q_J - \tilde{f}_{JK}P^K).
\end{eqnarray}
Using this ansatz the potential $V_b$ (\ref{eq:Vb}) can also be written 
\beq \label{black hole potential}
V_b= N^{IJ} \hat{Q}_I \bar{\hat{Q}}_J  + 2 e^K \,|Z|^2 
\;,
\end{equation}
where
\beq
\hat{Q}_I = Q_I - F_{IK} P^K  \hspace{8mm}  Z = P^I F_I - Q_I X^I \;.
\eeq

\end{document}